\documentclass[12pt]{article}
\usepackage{dcolumn}
\usepackage{xy}
\usepackage{graphicx}
\usepackage{epsfig}

\begin{document}
\def\1{\'{\i}}

\centerline{\Large\bf The Schr\"{o}dinger picture of standard
cosmology}
\vspace*{3mm}

\centerline{ N. Barbosa-Cendejas and M.A. Reyes}
\vspace*{2mm}

\centerline{\small\it Departamento de F\1sica, Universidad de
Guanajuato,\\ Apdo. Postal E143, 37150 Le\'on,Gto., M\'exico}
\vspace*{3mm}

\centerline{\bf \today}

\begin{center}
\small
\begin{minipage}{14cm}

We consider a time independent Schr\"{o}dinger type equation derived
from the equations of motion that drives a single scalar field in a
standard cosmology model for inflation in a flat space-time with a
Friedman-Robertson-Walker (FRW) metric with a cosmological constant.
We find that all the 1-dimensional bound state solutions of quantum
mechanics lead to at least one exact solution for the dynamical
equations of standard cosmology, and that these solutions resemble
the most recurrent inflationary solutions found in the literature.
The analogies derived from this approach may be used to realize a
deeper understanding of the dynamics of the model.


\vspace*{3mm}



\end{minipage}
\end{center}

\normalsize

\section{Introduction}

The current approach of inflation is that it occurs at some stage in
the early universe and that its source is one or several scalar
fields \cite{bassett}-\cite{Copeland}. The different models of
inflationary cosmology consider as a general feature a rapid growth
of the size of the universe at some stage in the early universe
\cite{cosmoliddle}-\cite{recinflpot}, this simple definition of
inflation may set the initial conditions for the large scale structure of the universe.
In this approach one
must consider an arbitrary functional form for the scalar field (SF)
potential $V(\phi)$, since there is no unique prescription or
phenomenology that could help to determine it.

Following this approach, let us consider a homogeneous and isotropic
Universe, i.e., a model in a FRW background with a scalar homogenous
field $\phi(t)$ minimally coupled to gravity and nonzero
cosmological constant
\begin{equation}
\int d^4x\sqrt{-g}\left[R+\Lambda+\frac{1}{2}(\nabla
\phi)^2+V(\phi) \right] , \label{action}
\end{equation}
where $(\nabla \phi)^2= g^{\mu\nu}\partial_{\mu}\partial_{\nu}$ and
$V(\phi)$ is the potential energy of the field. In order to describe
the dynamics of the scalar field during inflation the usual
treatment is performed \cite{bassett}-\cite{slowroll}, finally
leading to the pair of equations
\begin{eqnarray}
3 H^{2} &=&\frac{1}{2}\dot{\phi}^{2}+V(\phi)+\Lambda
\label{FRW1}
 \\
\ddot \phi+3H\dot\phi&=&-\frac{d V(\phi)}{d \phi},
\label{FRW}
\end{eqnarray}
where dot means derivative with respect to time, and we set $M_{Pl}=1$, $\hbar=c=1$.
The time derivative of eq.(\ref{FRW1}) is related to eq.(\ref{FRW}) through
the momentum equation
\begin{equation}
 \dot H=- \frac{1}{2} \dot{\phi}^{2}.
\label{Hdot}
\end{equation}

With the use of eqs. (\ref{FRW}) and (\ref{Hdot}) the
dynamics of the model may be described by the single equation:
\begin{equation}
3H^{2}+\dot{H}=V(\phi)+\Lambda. \label{Riccati}
\end{equation}
which can be recognized as a Riccati equation for the Hubble
parameter $H(t)$.  The Riccati equation has been one of the most
useful equations of mathematical physics, specially in
supersymmetric quantum mechanics (SUSY QM.) Its appearance here immediately
suggests a QM approach to inflationary cosmology based on the second
order differential equation derived from it. This has been proposed
earlier, the ansatz being to replace $x$ for $t$ and to assign
$a(t)=\psi(x)$, however leading to a nonlinear Schr\"odinger
equation \cite{nonlinearsch}. In this article we propose an
alternative transformation which leads to a linear Schr\"odinger
equation, where the SF potential $V(\phi)$ can be easily interpreted as
the QM potential $U(x)$, a simple scheme
whose consequences seem not to have been explored up to now. By
simple algebraic comparisons we end up with a powerful method to
derive particular exact solutions that may be useful for
understanding the inflationary period.
A similar approach for the case of cosmology with a perfect fluid has been proposed earlier,
but in the context of classical mechanics (\cite{Lima:1998kw,Lima:2001fi}.)
However, the fact that in our approach no restriction is made on the form of the SF potential
allows us to probe a deeper connection between QM and
inflationary cosmology.

\section{The Schr\"{o}dinger picture of Standard Cosmology }

\subsection{A Schr\"{o}dinger type equation for inflation }

By defining $\psi(t)$ through
\begin{equation}
H\equiv \frac{1}{3} \frac{\dot{\psi}(t)}{\psi(t)} ,
\label{hpsi}
\end{equation}
the Riccati equation (\ref{Riccati}) can be transformed into the one dimensional
Schr\"odinger equation
\begin{equation}
\left[ - \frac{d^{2}~}{dt^{2}}+3 ~ V(t)\right] \psi(t)= -3\Lambda~
\psi(t) \label{FRW-SCH},
\end{equation}
we shall consider solutions to eq.(\ref{FRW-SCH}) based only on the
fact that the Hubble parameter $H(t)$ cannot be a singular function,
implying that $\psi(t)$ has to be an at least $C^{1}$ class function
without zeros, but without any other restriction. For example, we
may consider all ground state solutions of known exactly solvable
bound state problems in QM as solutions to eq.(\ref{FRW-SCH}).
Hence, an immediate equivalence arises between the SF potential
$V(\phi)$ and the cosmological constant $\Lambda$, with the QM
potential $U(x)$ and ground state energy eigenvalue $E_g$,\footnote{ For simplicity
we shall use $m=1$ for the Schr\"odinger particle.} respectively,
\begin{equation}\label{v.eq.u}
3V(\phi(t))+ 3\Lambda \leftrightarrow 2 U(x)-2 E_{g}
\label{vlue}
\end{equation}
This is indeed a very simple proposal, which shows that all the
known exactly solvable stationary problems of 1-dimensional QM must
provide at least one exact solution to the cosmological
Schr\"odinger type equation. The general algebraic procedure is very simple:
for any given QM problem, use the ground state eigenfunction $\psi_{g}(x)$ and eq.(\ref{hpsi})
to find $H(t)$; then, use eq.(\ref{Hdot}) to find $\phi(t)$, which together with eq.(\ref{v.eq.u})
defines $V(\phi)$.

For example, in the QM case of the simple harmonic oscillator (SHO), where $U(x)=\omega^2x^2/2$,
the Schr\"odinger equation
\begin{equation}
\left[-\frac{d^{2}}{dx^{2}}+2 \left( \frac{\omega^2}{2}x^2-E_{n}
\right) \right ]\psi_{n}(x)=0 .
\end{equation}
has the wave functions and energy eigenvalues given by
\begin{equation}
\psi_{n}(x)= \sqrt{\frac{1}{2^n n!}\sqrt{\frac{\omega}{\pi}}} ~
e^{-\frac{w}{2}x^2}H_{n}(\sqrt{\omega}\, x)~~~~~~~
E_{n}=\left(n+\frac{1}{2} \right)\omega
\end{equation}
where $H_{n}(y)$ are the Hermite polynomials.  In the case $n=0$,
the Hermite polynomial is $H_{0}(\sqrt{\omega}\, x)=1$, with energy and
wave function
\begin{equation}
E_{0}= \frac{1}{2}\omega~~~~~~~~~
\psi_{0}(x)=\sqrt[4]{\frac{\omega}{\pi}} ~ e^{-\frac{\omega}{2}x^2} \ .
\label{E0sho}
\end{equation}
To construct the corresponding cosmological variables, we replace $x$
by $t$ in eq.(\ref{E0sho}), and use eq.(\ref{hpsi}) to find the
associated Hubble's parameter
\begin{eqnarray}
H(t)=-\frac{\omega}{3}t
\end{eqnarray}
and with the use of eq.(\ref{Hdot}) we obtain the expression of the
scalar field
\begin{equation}
\phi(t) = \sqrt{\frac{2w}{3}}~t \label{phisho}
\end{equation}
Finally, we use eqs.(\ref{vlue}) and (\ref{phisho}) to find the SF potential $V(\phi)$ and the constant $\Lambda$,
\begin{equation}
V(\phi)=\lambda\phi^2
\ , \ \ \ \
\Lambda=-\frac{2}{3}\lambda
\end{equation}
where $\lambda=\frac{\omega}{2}$. As one can see, the scalar field
potential derived from the SHO potential turns out to be
$\lambda\phi^2$ . Surprisingly, one of the most
useful and basic potentials of QM transforms into one of the most
useful potentials in this cosmological model (see \cite{bassett},
\cite{luisu}, and references there in.)  It is even more surprising
that other typical QM potentials resemble typical scalar field
potentials in standard cosmology; for example, compare the results
in \cite{Kiselev} and \cite{tonatiuh} to the ones obtained in Table
1.

\subsection{Inflationary solutions from QM central force problems}

In Table 1 we have left some free parameters which can be used to fix the cosmological variables strength, even though their overall behavior is already
determined.  For example, the cosmological solutions above may be set to correspond to the half plane solutions of the Schr\"odinger problem, with the initial condition $a(0)\ne 0$ if $t=0$ when $x=0$,
and it is not possible to have as initial condition $a=0$.
Since in the half plane not only the QM ground
state function is nodeless, but also the first excited state
function is, we end up with two exact standard
cosmology solutions with different initial conditions derived from
each 1-dimensional QM problem in the half plane $t\ge 0$,
the second one with the initial
condition $a(0)=0$.

Following this discussion, it is obvious that all ground state solutions to the radial problem
\begin{equation}
-\frac{d^2u_{nl}}{dr^2} + 2\left[ U(r)+\frac{l(l+1)}{2 r^2}
-E_{nl}\right] u_{nl} = 0
\end{equation}
are useful to derive exact solutions to the SF equations. Moreover,
since all eigenfunctions of the radial problems with angular
momentum $l=n-1$ are nodeless, we end up with an infinity
of exact solutions to the SF dynamics equations, paratemeterized by a cosmological constant
that belongs to the discrete set  $\Lambda_n=-\frac{1}{3}E_n$, with the SF potential
always including the centrifugal barrier term $n(n-1)/t^2$, and with the initial condition $a(0)=0$.

As an example, let us
consider the Hydrogen like potential $U(r)=-Zq^2/r$ were all the
nodeless eigenfunctions are given by
\[
u_{n,n-1}(r) \propto r^n e^{-Z\alpha\, r/n} \ .
\]
where the fine structure constant $\alpha$  and the atom
number $Z$, are constants only used to determine the strength of the
cosmological variables. For the ground state ($n=1$) the cosmological variables are
\begin{eqnarray}
H(t)&=& \frac{1}{3t}-\frac{Z\alpha}{3} \nonumber  \\
\phi(t)&=& \sqrt{\frac{2}{3}}\ln t \\
V(\phi)&=& -\lambda e^{-\sqrt{3/2}\phi} \nonumber \\
\Lambda&=&\frac{\lambda Z}{2} \nonumber
\end{eqnarray}
where $\lambda=2Z\alpha^2/3$, while for $n>1$, they become
\begin{eqnarray}
H(t)&=& \frac{n}{3t}-\frac{Z\alpha}{3n} \nonumber \\
\phi(t)&=& \sqrt{\frac{2n}{3}}\ln t \\
V(\phi)&=& \frac{n(n-1)}{3} \left[e^{2\sqrt{3/2n}\,\phi}
-\frac{2Z\alpha}{n(n-1)}
e^{-\sqrt{3/2n}\, \phi} \right]  \nonumber \\
 \Lambda&=& \frac{Z^2\alpha^2}{3n^2} \nonumber
\end{eqnarray}
Hence, for $n=1$ the SF potential becomes an exponential potential,
while for $n>1$, if we choose $Z\alpha=n(n-1)$ it becomes a Morse potential.
The appearance of this Morse
potential is a very interesting feature of this model, since this
potential is very slowly varying for $t\to \pm \infty$, has a very
soft minimum, and later exponentially grows for $t \to \mp \infty$,
the sign depending on the parameters. Therefore, this potential has
all the desired features to allow a SF $\phi(t)$ slowly roll to the
minimum of the SF potential $V(\phi)$ \cite{marcoyotros}. All these
results are also included in Table 1.

\begin{table}
\small
\begin{center}
\begin{tabular}{|l|c|c|}
  \hline
  \multicolumn{2}{|c|}{Schr\"{o}dinger picture} &
  \multicolumn{1}{c|}{Standard Cosmology} \\ \hline
  ~ & ~ & ~ \vspace*{-3mm} \\

  ~ & $\psi_0(x)= \sqrt[4]{\frac{\omega}{\pi}} ~ e^{-\frac{\omega}{2}x^2}$
  & $H(t)=-\frac{2\lambda}{3}\, t$,~~$\phi(t)=2\sqrt{\frac{\lambda}{3}}\,t$   \\
  ~ & ~ & ~ \vspace*{-4mm} \\
  SHO &  $E_0=\frac{1}{2}\omega$ & $\Lambda=-\frac{2}{3}\lambda$ \\
  ~ & ~ & ~ \vspace*{-4mm} \\
  ~ & $U(x)=\frac{1}{2}\omega^2 x^2$ & $V(\phi)=\lambda\phi^2$ \ , \ \ \ $\lambda=\frac{\omega}{2}$ \\

~ & ~ & ~ \vspace*{-3mm} \\
\hline
~ & ~ & ~ \vspace*{-3mm} \\

  ~ & $\psi_0(x)= 8e^{-2e^{-\alpha x}}e^{-\frac{3}{2}\alpha x} $
  & $H(t)=\frac{32}{3}\sqrt{\frac{\lambda}{3}} \left( e^{-16\sqrt{\frac{\lambda}{3}}\, t}-\frac{3}{4} \right)$,~~
  $\phi(t)=-\frac{4}{\sqrt{3}}e^{-4\sqrt{\frac{\lambda}{3}}\, t}$  \\
  ~ & ~ & ~ \vspace*{-4mm} \\
  MOR &  $E_0=-\frac{9}{8} \alpha^2$
  & $\Lambda=16\lambda$ \\
  ~ & ~ & ~ \vspace*{-4mm} \\
  ~ & $U(x)=2\alpha^2 (e^{-2 \alpha x}-2 e^{-\alpha x})$
  & $V(\phi)=\lambda\phi^{4}-\frac{32}{3}\lambda\phi^{2}$ \ , \ \ \ $\lambda=\frac{3\alpha^2}{64}$ \\

~ & ~ & ~ \vspace*{-3mm} \\
\hline
~ & ~ & ~ \vspace*{-3mm} \\

  ~ & $\psi_0(x)= \frac{1}{2^{\lambda}}
  \cos^{\lambda}(\alpha x)$
  & $H(t)=-\frac{\alpha\lambda}{3} \tan(\alpha t)$,~~$\phi(t)=\sqrt{\frac{2\lambda}{3}}
~\ln\sqrt{\frac{1+\sin(\alpha t)}{1-\sin(\alpha t)}}$  \\
  ~ & ~ & ~ \vspace*{-4mm} \\
  PTT &  $E_0=\frac{\alpha^2\lambda^2}{2}$
  & $\Lambda=-\frac{\lambda^2\alpha^2}{3}$ \\
  ~ & ~ & ~ \vspace*{-4mm} \\
  ~ & $U(x)=\frac{\alpha^2}{2}\frac{\lambda(\lambda-1)}{\cos^2(\alpha x)}$
  & $V(\phi)=\frac{\alpha^2}{3}\lambda(\lambda-1)
  \cosh^2\!\left(\sqrt{\frac{3}{2 \lambda}} \phi\right)$ \\

~ & ~ & ~ \vspace*{-3mm} \\
\hline
~ & ~ & ~ \vspace*{-3mm} \\

  ~ & $\psi_0(x)= \frac{1}{\cosh{\alpha x}}$
  & $H(t)=-\frac{\alpha}{3} \tanh(\alpha t)$,~~$\phi(t)=\sqrt{\frac{2}{3}}\arcsin(\tanh(\alpha t))$  \\
  ~ & ~ & ~ \vspace*{-4mm} \\
  PTH &  $E_0=-\frac{\alpha^2}{2}$
  & $\Lambda=\frac{\alpha^2}{3}$ \\
  ~ & ~ & ~ \vspace*{-4mm} \\
  ~ & $U(x)=\frac{- \alpha^2}{\cosh^2(\alpha x)}$
  & $V(\phi)=-\frac{2\alpha^2}{3}\cos^2\!\left(\sqrt{\frac{3}{2}}\phi\right)$ \\

~ & ~ & ~ \vspace*{-3mm} \\
\hline
~ & ~ & ~ \vspace*{-3mm} \\

%

  ~ & $u_{n,n-1}(r)= \frac{(2\alpha)^{3/2}}{\sqrt{2n^4(2n-1)! }}
  \left(\frac{2\alpha r}{n}\right)^n e^{-\alpha r/n}$
  & $H(t)=\frac{n}{3t}-\frac{\alpha}{3n}$,~~ $\phi(t)=\sqrt{\frac{2n}{3}}\ln t$  \\
  ~ & ~ & ~ \vspace*{-4mm} \\
  HA &  $E_{n}=-\frac{\alpha^2}{2n^2}$
  & $\Lambda=\frac{\alpha^2}{3n^2}$ \\
  ~ & ~ & ~ \vspace*{-4mm} \\
  ~ & $U_{{e\!f\!f}}(r)= -\frac{\alpha}{r} + \frac{n(n-1)}{2r^2}$
  & $V(\phi)=
\frac{n(n-1)}{3} \left[ e^{-2\sqrt{\frac{3}{2n}}\, \phi}-
\frac{2\alpha}{n(n-1)} e^{-\sqrt{\frac{3}{2n}}\, \phi} \right]$ \\

\hline

\end{tabular}
\caption{Standard Cosmology exact solutions from six Schr\"odinger
problems: simple harmonic oscillator (SHO), Morse potential (MOR),
Trigonometric P\"oschl Teller (PTT), Hyperbolic P\"oschl Teller
(PTH), and Hydrogen Atom (HA).}
\end{center}
\label{tabl1}
\end{table}
\normalsize

\subsection{QM and Standard Cosmology analogies}

Looking at Table 1, the Schr\"odinger picture of standard cosmology
seems to be a fruitful approach to the construction of exact solutions to the
inflationary equations (\ref{FRW1},\ref{FRW}), since all potentials from these known
QM problems resemble known SF potentials. It may seem that there must exist
further analogies between these two models of the micro and macro
cosmos than just an algebraic resemblance.

In the present analogy $\psi(t) = a^3(t)$ describes the way the
universe volume is expanding since in the Schr\"odinger picture,
$\psi(x)$ is related to probability conservation, hence the
equivalence proposed here points to energy density conservation in
this expanding universe. On the other hand, the only constant term
in the QM problem is the energy $E$, which therefore determines the
cosmological constant $\Lambda$ of eq.(\ref{FRW-SCH}), which could
be associated with the vacuum energy density. Therefore, the sign of
$\Lambda$ is completely determined by the corresponding QM problem
from which the SF solution is derived, becoming an immediate check
for the dynamical characteristics that one wants to determine with
the proposed SF potential.

With respect to the scalar field $\phi(t)$, following eqs.(\ref{Hdot}), (\ref{Riccati}) and
(\ref{v.eq.u}), we can see that wherever $\dot a(t)\simeq 0$,
\begin{equation}
\phi(t) \simeq  \int^t dy\sqrt{2\left(E-U(y)\right)
}
\end{equation}
which resembles the action $S(x)$ of the quantum theory.

\vspace*{1cm}


\subsection{Slow Roll and WKB approximations}

One further analogy deserves special attention.  Beginning with the slow roll approximation condition
\[
\left| \frac{V'}{V}\right|^2 < 1
\label{slowrll}
\]
where we should do the substitution $V \to V+\Lambda$ to comply with  eq.(\ref{Riccati}), we can see that since $\ddot a/a>0$ implies that $V+\Lambda>\dot\phi ^2$, we have that
\[
\left| \dot V \right|^2= \left| \dot \phi V'  \right|^2 < \left| V+\Lambda \right| \left| V' \right|^2
< \left| V+\Lambda \right|^3
\]
In our QM analogy, we would have to do the substitutions $\frac{dV}{dt} \to \frac{dU}{dx}$ and
$\left| V+\Lambda \right| \to \left| E-U \right| $, giving
\[
\left| \frac{dU}{dx} \right|^2 < \left| E-U \right|^3
\]
which is just the WKB approximation
\[
\left| \frac{d^2W}{dx^2} \right| < \left| \frac{dW}{dx} \right|^2
\]
of the stationary problem, where $W(x)=\pm \sqrt{2(E-U)}$ is Hamilton's principal function.


\section{An ever expanding universe}
All the QM problems considered here lead to two possible initial
conditions for the scale factor, $a(0)=0$ or $a(0)> 0$, but with only one
final condition, $a(t)\to 0$ as $t\to\infty$, if eqs.(\ref{FRW1},\ref{FRW})
may be used for all the half plane $t\ge 0$.  If this Big Crunch could
not be attainable, as observations seem to predict, our simple approach
may still be useful to describe the expected dynamics.

If $a(t)$ is always increasing, then a QM bound state is not the
right solution.  However, as is depicted in Ref.\cite{tesismarco},
the Schr\"odinger equation (\ref{FRW-SCH}) has an infinite number of
wave functions that diverge to $\pm \infty$ as $t\to \infty$, for
the continuum set of energies $E$, with $E_n< E <E_{n+1}$,  as is
depicted in Fig.(1) for the case of the SHO.

\begin{figure}[h]
 \begin{center}
 \epsfig{figure=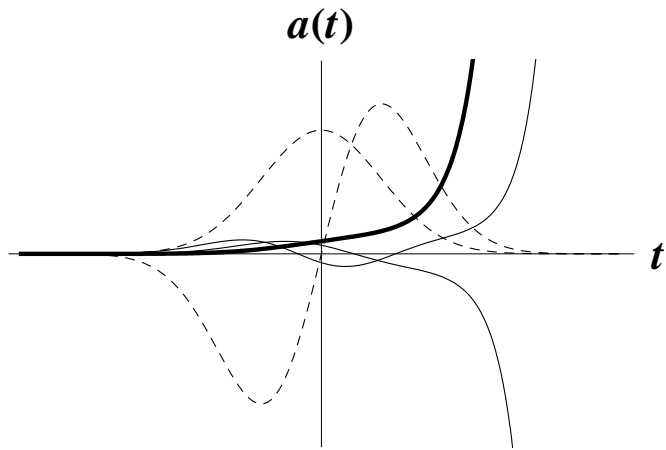, height=6cm}
 \end{center}
 \caption{Different forms of $a(t)$ obtained from one QM problem.}
\end{figure}

In Fig.(1) the dashed curves correspond to $a(t)$ obtained from the
ground state and first excited state eigenfunctions, $\psi_0(x)$ and
$\psi_1(x)$. These two curves have $a(t)\to 0$ as $t\to\infty$. On
the other hand, the solid curves draw the scale factor for three
different cosmological constants, derived from the energy eigenvalues,
$E\!<\!E_0$, $E_0\!<\!E\!<\!E_1$ and
$E_1\!<\!E\!<E_2$, whose wave functions diverge to $+\infty$,
$-\infty$ and $+\infty$ again, the first one without nodes and the
other with increasing number of nodes. Therefore, only the wave
function with $E\!<\!E_0$ could lead to a physical solution
$a(t)\!>\!0$ for all $t$, describing a cosmological solution for an ever
expanding universe.

\section{Conclusion}

In this article we present a simple ansatz that links two completely
separate models, one being a cosmological model for inflation of the
macrocosmos, and the other a quantum model for the microcosmos.
Surprisingly, we have found that the most common potential functions
in these two models map from one to the other, with probability
conservation in QM reflecting into energy density conservation in
the cosmological model, a cosmological constant determined by
the energy eigenvalue in the former, and the WKB approximation reflecting into
the slow roll approximation.  All these analogies define
deep connection between the two models, with respect to their
dynamical behaviors. Finally, our initial proposal was to look at
the bound state solutions of the QM problems to describe a non
singular Hubble variable for the cosmological model, but since
present observations seem to indicate that the universe is not
receding but on the contrary accelerates with time, we may have to
recur to the unbounded solutions of the Schr\"odinger problem,
supporting even more the validity of the simple analogy developed in
this work.


\section{Acknowledgements}

We would like to thank M. Sabido for useful comments.
We also ackwoledge support from CONACYT of Mexico, through a scholarship for NBC, and
through grant No. SEP-2003-C0245364.


\end{document}